\def\etcl{$\kappa$-(BE\-DT\--TTF)$_2$\-Cu\-[N\-(CN)$_{2}$]Cl}
\def\etcn{$\kappa$-(BE\-DT\--TTF)$_2$\-Cu$_2$(CN)$_{3}$}

\def\khgcl{$\kappa$-(BE\-DT\--TTF)$_2$\-Hg\-(SCN)$_{2}$\-Cl}
\def\bhgcl{$\beta^{\prime\prime}$-(BE\-DT\--TTF)$_2$\-Hg\-(SCN)$_{2}$\-Cl}
\def\dhgcl{$\beta^{\prime\prime}$-(D$_8$-BE\-DT\--TTF)$_2$\-Hg\-(SCN)$_{2}$\-Cl}
\def\cm{cm$^{-1}$}

\documentclass[pra,twocolumn,superscriptaddress,amsmath,amssymb]{revtex4}
\usepackage{graphicx}
\usepackage{units}
\usepackage{color}
\usepackage[dvipsnames]{xcolor}
\usepackage{eqnarray}
\usepackage{tabularx}
\usepackage{bbm}
\usepackage{changes}
\usepackage{textcomp}

\def\cm{cm$^{-1}$}
\def\be{\begin{equation}}
\def\ee{\end{equation}}
\def\ba{\begin{eqnarray}}
\def\ea{\end{eqnarray}}

\begin{document}
\title{The metal-insulator transition in the organic conductor $\beta^{\prime\prime}$-(BEDT-TTF)$_{2}$Hg(SCN)$_{2}$Cl}
    \author{Weiwu Li}
    \affiliation{1.~Physikalisches Institut, Universit\"at Stuttgart, Pfaffenwaldring 57, 70550 Stuttgart, Germany}    \author{Eva Rose}
    \affiliation{1.~Physikalisches Institut, Universit\"at Stuttgart, Pfaffenwaldring 57, 70550 Stuttgart, Germany}        \author{Minh Vu Tran}
    \affiliation{1.~Physikalisches Institut, Universit\"at Stuttgart, Pfaffenwaldring 57, 70550 Stuttgart, Germany}    \author{Ralph H\"{u}ber}
    %\author{R. R\"osslhuber}
    \affiliation{1.~Physikalisches Institut, Universit\"at Stuttgart, Pfaffenwaldring 57, 70550 Stuttgart, Germany}
    \author{Andrzej \L{}api\'{n}ski}
    \affiliation{Institute of Molecular Physics, Polish Academy of Sciences, Smoluchowskiego 17, 60-179, Pozna\'{n}, Poland}
    \author{Roman \'{S}wietlik}
    \affiliation{Institute of Molecular Physics, Polish Academy of Sciences, Smoluchowskiego 17, 60-179, Pozna\'{n}, Poland}
    \author{Svetlana A. Torunova}
    \affiliation{Institute of Problems of Chemical Physics, Russian Academy of Sciences, 142 432, Chernogolovka, Russia}
    \author{Elena I. Zhilyaeva}
    \affiliation{Institute of Problems of Chemical Physics, Russian Academy of Sciences, 142 432, Chernogolovka, Russia}
    \author{Rimma N. Lyubovskaya}
    \affiliation{Institute of Problems of Chemical Physics, Russian Academy of Sciences, 142 432, Chernogolovka,
Russia}
    \author{Martin Dressel}
    \affiliation{1.~Physikalisches Institut, Universit\"at Stuttgart, Pfaffenwaldring 57, 70550 Stuttgart, Germany}
    \date{\today}

\begin{abstract}
We explore the nature of the metal-insulator transition in the two-dimensional organic compound $\beta^{\prime\prime}$-(BEDT-TTF)$_{2}$Hg(SCN)$_{2}$Cl by x-ray, electrical transport, ESR,
Raman and infrared investigations.
Magnetic and vibrational spectroscopy concurrently reveal a gradual dimerization along the stacking direction $a$, setting in already at the crossover temperature of 150~K from the metallic to the insulating state. A spin gap $\Delta_\sigma = 47$~meV is extracted.
From the activated resistivity behavior  below $T=55$~K a charge gap $\Delta_\rho=60$~meV is derived.
At $T_{\rm CO} = 72$~K the C=C vibrational modes reveal the development of a charge-ordered state with a charge disproportionation of $2\delta_{\rho}=0.34e$. In addition to a slight structural dimerization, charge-order causes horizontal stripes perpendicular to the stacks.
\end{abstract}
\maketitle

\section{Introduction}

Quasi two-dimensional charge-transfer salts based on the organic molecule BEDT-TTF (where BEDT-TTF stands for bis-ethylenedithio-tetrathiafulvalene) provide a unique playground for studying the interaction of electronic and structural properties because they exhibit a rich phase diagram ranging from metals and superconductors to charge-ordered and Mott insulators, often combined with magnetic order but also spin-liquid properties \cite{ToyotaBook,LebedBook,MoriBook}.
These compounds attract great attention in experimental and theoretical condensed-matter physics because their behavior can be tuned by chemical substitution or external pressure; thus certain properties can be strengthened or suppressed as desired. The wealth of possibilities gets even enlarged by the fact that many of these compounds possess several polymorphic structures (labelled by Greek letters) often with quite different properties. Due to the D$_2$A stoichiometry (with D the organic donor and A the inorganic acceptor components), the conduction bands are supposed to be three-quarter filled, leading to charge-ordered ground states if the intersite Coulomb repulsion is strong enough, as observed in numerous $\alpha$- \cite{Wojciechowski03,Dressel03,Drichko06,Yue10,Ivek11}, $\beta^{\prime\prime}$- \cite{Yamamoto04,Yamamoto06,Yamamoto08,Kaiser10} and $\theta$-compounds \cite{HMori98}, for instance. In dimerized structures, such as the $\kappa$-phase, however, the bands are half-filled, with Mott physics playing the superior role \cite{ToyotaBook,LebedBook} and typically no tendency to charge order \cite{Sedlmeier12}.

%Recently this interplay of onsite and intersite Coulomb repulsion moved in the focus of research when the dimerized compound \khgcl\ with a half-filled conduction band was shown to develop charge order at low temperatures \cite{Drichko14},while other Mott insulators, such as \etcl, \etcn\ or $\kappa$-(BE\-DT\--TTF)$_2$\-Ag$_2$(CN)$_{3}$,do not exhibit any sign of charge disproportionation \cite{Sedlmeier12,Pinteric16}.In order to shed light on this issue, we have performed pressure-dependent investigations on \khgcl\ \cite{Lohle17} and substituted Br for Cl \cite{Ivek17}. Here we consider structural variations.

Among those weakly dimerized phases, the non-mag\-ne\-tic $\beta^{\prime\prime}$-compounds are of particular interest as several species exhibit superconductivity at ambient condition or under pressure \cite{HMori98}. Merino and McKenzie \cite{Merino01} predicted that a superconducting state could be realized by charge fluctuations
when the long-range charge order is suppressed that commonly leads to an insulating state.
The underlying principle differs from the superconducting mechanism found in the $\kappa$-phase
where magnetic fluctuations are of superior importance.
Their suggestion was confirmed by different experimental methods such as, infrared spectroscopy \cite{Kaiser10} and NMR \cite{Kawamoto11}. For a better understanding of the superconducting mechanism, it is important to study also the charge-order state above $T_c$.
Previously comprehensive optical investigations have been conducted \cite{Yamamoto04,Yamamoto06,Yamamoto08,Kaiser10,Girlando14} on several $\beta^{\prime\prime}$-salts with different correlating strength.
Here we want to extend the efforts to the charge-order state in a new $\beta^{\prime\prime}$-salts, which we mainly explore by optical spectroscopy in order to understand the phase diagram of these compounds.

Already in 1993 Lyubovskaya and collaborators \cite{Lyubovskaya93,Aldoshina93}
synthesized the family of salts with the composition (BEDT-TTF)$_2$Hg(SCN)$_{3-n}$$X_n$ ($X$ = F, Cl, Br and I and n=1 or 2) that has drawn increasing attention recently. Among them \khgcl\ is of particular interest as the compound exhibits sharp metal-insulator transition around $T_{\rm CO}=30$~K without structural changes \cite{Yasin12,Drichko14,Lohle17}.
It was also reported \cite{Lyubovskaya95} that in the process of synthesis and crystal growth a new salt of the same chemical composition but different structure and properties was obtained. The present paper is devoted to the complete
characterization of \bhgcl, using x-ray analysis, infrared and Raman spectroscopy, transport and ESR measurements.

\section{Experiment Method}
Single crystals of $\beta^{\prime\prime}$-(BEDT-TTF)$_{2}$Hg(SCN)$_{2}$Cl were prepared by electrochemical oxidation of the BEDT-TTF molecules.% which stands for bis-ethylenedithio-tetra\-thia\-fulvalene.
These crystals are formed in the synthesis along with $\kappa$-(BEDT-TTF)$_2$Hg(SCN)$_2$Cl crystals \cite{Lyubovskaya93,Aldoshina93}. Typically, a 4~ml solution of BEDT-TTF (10~mg, 0.026~mmol) in 1,1,2-trichloroethane (TCE) was added to the anode compartment of the cell, and a 10~ml solution containing Hg(SCN)$_2$ (34.2~mg, 0.108~mmol), [Me$_4$N]SCN$\cdot$KCl (14.3~mg, 0.069~mmol), and dibenzo-18-crown-6 (32.1~mg, 0.09~mmol) in 12\%\ ethanol/TCE was added to the cathode compartment and the anode compartment, to level both sides. Electrocrystallization was carried out at a temperature of $40^\circ$~C and a constant current of 0.5$\mu$A. The crystals of $\beta^{\prime\prime}$- and $\kappa$-phases were unambiguously identified by electron spin resonance (ESR) spectroscopy ($\Delta H_{pp} = 25-38$~G and 60-90~G, respectively) or by temperature of a transition to the insulating state.
The resulting single crystals of the $\beta^{\prime\prime}$-phase are platelets with a typical size  of $1 \times 1 \times 0.04$~mm$^{3}$; the larger face corresponds to the highly conducting ($ab$)-plane.
X-ray diffraction data were collected at $T=293$~K using a Bruker Kappa Apex2duo diffractometer with Mo-K$_{\alpha}$ radiation ($\lambda = 0.71073$~{\AA}). The structure was solved by direct method and refined by full-matrix least-squares techniques on $F^2$ employing the program system SHELX-97 \cite{Sheldrick08,remark1}.

The temperature-dependent dc-resistivity was measured by standard four-point technique in a custom-made helium bath cryostat with a cooling and warming rate of less than 0.3~K/min. For the current and voltage contacts $15~\mu$m golden wires were glued directly to the sample with carbon paste; the voltage was fixed to 2~mV.

Electron-spin-resonance (ESR) measurements were carried out by utilizing a continuous-wave X-band spectrometer (Bruker ESR 300) equipped with an Oxford ESR 900 cryostat for temperature-dependent measurements between $T=10$ and 240~K.
The sample was glued to a quartz rod by vacuum grease in such a way that the external magnetic field is oriented within the conducting plane. The signal could be fitted with Lorentzians;
relative paramagnetic susceptibility was estimated from the line intensity using $\chi\propto(\Delta H_{\rm pp})^{2}I_{\rm max}$, where $ \Delta H_{\rm pp}$ is the peak-to-peak field and $I_{\rm max}$ the intensity of the ESR signal.

The Raman spectra of \bhgcl\ single crystals were recorded in backscattering geometry
within the energy range of $1000-1500~{\rm cm}^{-1}$ with 1~\cm\ spectral resolution using a Horiba Jobin Yvon Labram HR 800 spectrometer equipped with a liquid-nitrogen-cooled CCD detector. The incident He-Ne laser beam ($\lambda = 632.8$~nm) was focused on the ($ab$)-plane of the salt, and the scattered beam was collected without polarization analysis. Temperature-dependent spectra were obtained by mounting the specimen on a continuous-flow helium cryostat and cooling down with a typical rate of 1~K/min.

Polarized infrared reflectivity measurements were performed over a broad energy range (from 200 to 8000~\cm, 1~\cm\ resolution) from room temperature down to 10~K using a Brucker Vertex 80v Fourier-trans\-form infrared spectrometer equipped with a microscope (Bruker Hyperion) and a CryoVac microcrystat. The absolute value of reflectivity was obtained by comparison to an aluminum mirror. The optical conductivity was extracted from the reflectivity data via Kramers-Kroning transformation. At low frequencies, the data were extrapolated by the Hagen-Rubens relation for the metallic state and a constant value for the insulator state, while the high-energy part was extended up to 50\,0000~\cm\ using a free-electron model $R(\omega) \propto \omega^{-4}$.

\section{Results and Discussion}
\subsection{Crystal structure}
\begin{figure}
	\centering
	\includegraphics[width=\columnwidth]{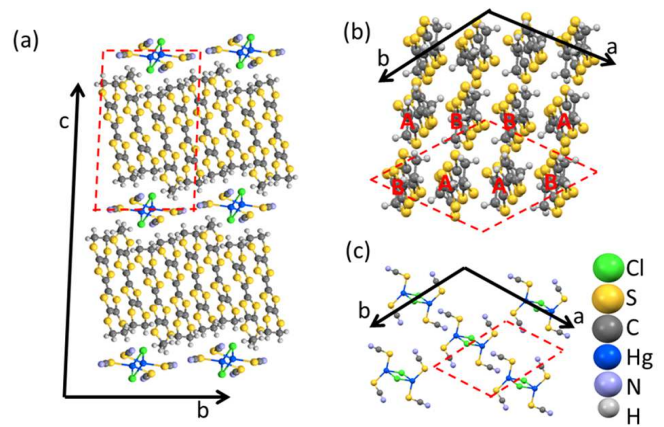}
	\caption{(Color online) Crystal structure of  $\beta^{\prime\prime}$-(BEDT-TTF)$_{2}$Hg(SCN)$_{2}$Cl at room temperature.Unit cell borders are marked with red dashed lines (a)~Projection of the molecular arrangement along the $a$-axis illustrates the alternating cation and anion layers. (b)~Packing pattern of the BEDT-TTF molecules in the donor layer with two distinct molecules, hereafter designated as A and B. The A\,B\,B\,A\,A\,B\,B\,A-stacks are basically form along the $(a-b$)-direction. (c)~Dimer arrangement of the Hg$_2$(SCN)$_{4}$Cl$_2$ unit in the anionic layer.}
	\label{fig:structure}
\end{figure}
The unit-cell parameters of \bhgcl\ at room-temperature are summarized in Table~\ref{tab}.
\begin{table}[b]
\caption{Crystallographic data for $\beta^{\prime\prime}$-(BEDT-TTF)$_{2}$\-Hg(SCN)$_{2}$Cl obtained from x-ray diffraction ($\lambda = 0.71073$~\AA) at ambient conditions ($T=293$~K) \cite{remark1}. For comparison we also list the data for the deuterated analogue \cite{Dyachenko95}.}
\begin{tabular}{lll}
		Chemical formula  & C$ _{22}$H$_{16}$S$_{18}$N$_{2}$HgCl~~ & C$ _{22}$D$_{16}$S$_{18}$N$_{2}$HgCl     \\
		Form. weight $M_{W}$           &  1121.49 &    1137.59   \\
		Crystal system & triclinic & triclinic\\
		Space group    &  P$\overline{1}$  & P$\overline{1}$ \\
		$a$ (\AA{})       &	9.568(2)        & 9.717(3) \\
		$b$ (\AA{})       &	10.778(2)         & 11.067(3) \\
		$c$ (\AA{})       &18.844(4)       	& 19.348(5)  \\
		$\alpha$ (deg.)   &	90.60(3)   &   77.69(2)  \\
		$\beta$ (deg.)    &102.13(3) & 106.90(2)\\
		$\gamma$ (deg.)   &	113.66(3)& 114.28(3) \\
		Volume $V$ (\AA$^{3}$)   & 1730.4(6)	& 1804.2(7)\\
		$Z$	              & 2	& 2 \\
		Density $D_{c}$ (g/cm$^{3}$)	& 2.152	& 2.11 \\
		Absorption coeff. & &	\\
        ~~~$\mu$ (mm$^{-1}$)	&  5.635 & 5.4 \\
		F(000)&	1094 & 1094 \\
No.\ of refl.\ meas.\ & 41480 & \\ 		
No.\ of indep.\ refl.\ & 9989 & 3784 \\
$R_1$ & 0.0216 & 0.05 \\
$\omega R_2$ & 0.0556 & 0.05 \\
GOF & 0940 & \\
	\end{tabular}
	\label{tab}
\end{table}
A deuterated sister compound (D$_8$-BEDT-TTF)$_{4}$[Hg(SCN)$_{2}$Cl]$_2$ with a $\beta$-type packing was prepared previously  \cite{Dyachenko95,Lyubovskaya95,Yudanova94}.
As common for these charge transfer salts, layers of BEDT-TTF radical cations and Hg(SCN)$_{2}$Cl anions are alternatingly stacked along the crystallographic $c$-axis as shown in Fig.~\ref{fig:structure}(a).
Within the cationic layers, the arrangement of the BEDT-TTF molecules can be best described by the $\beta^{\prime\prime}_{412}$-type packing motif \cite{TMori98} with tilted stacks along the ($a$-$b$)-axis. As illustrated in
Fig.~\ref{fig:structure}(b) two crystallographically non-equivalent BEDT-TTF molecules (labelled A and B) can be distinguished with planar and nonplanar structures of the TTF fragments, respectively. The distances between (A,A) and (B,B) are similar but slightly longer than that of (A,B), causing a weak structural dimerization. Since neutral TTF-fragments are commonly bent while planar TTF-fragments are charged,
this might suggest that in the present case the charge is distributed non-uniformly. The anionic layers
consist of doubly charged [Hg$_{2}$(SCN)$_{4}$Cl$ _{2} $]$^{2-}$  dimers as shown in Fig.~\ref{fig:structure}(c).
This is distinctively different from the single-charged monomeric anions [Hg(SCN)$_{2}$Cl]$^{-} $ present in
the $\kappa$-phase analogue \cite{Drichko14}.
%For that reason the nomenclature $\beta^{\prime\prime}$-(BEDT-TTF)$_{4}$[Hg(SCN)$_{2}$Cl]$_2$ suggested in Ref.~\onlinecite{Yudanova94,Dyachenko95,Lyubovskaya95} is more appropriate.

\subsection{Transport Properties}
\begin{figure}
	\centering
	\includegraphics[width=\columnwidth]{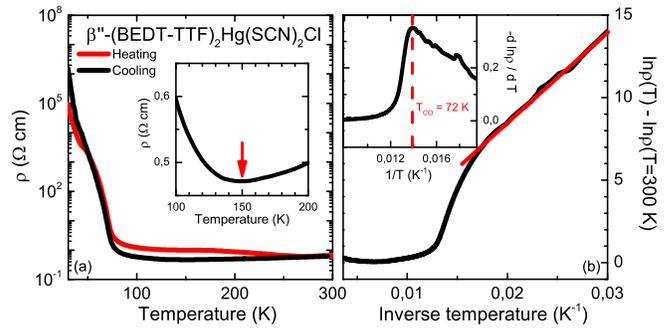}
	\caption{(Color online) (a)~Temperature dependence of the dc resistivity measured in the conducting ($ab$)-plane of \bhgcl. The inset magnifies the minimum in $\rho(T)$ around $T_{m}=150$~K. (b)~The resistivity plotted as a function of inverse temperature $1/T$. The red straight line indicates an activation energy of $\Delta_\rho = 60$~meV, which corresponds to 700~K.
The logarithmic resistivity derivative is plotted in the inset versus the inverse temperature; the dashed line indicates the charge order transition $T_{CO}=72$~K. }
	\label{fig:dc}
\end{figure}

Fig.~\ref{fig:dc}(a) presents the temperature dependence of the electrical resistivity $\rho(T)$ of \bhgcl\ measured within the highly conducting ($ab)$-plane. At room temperature the conductivity is around $2~(\Omega{\rm cm})^{-1}$, which is a typical value for organic conductors.
From $T=300$ down to 150~K this compound shows a weakly metallic behavior, before the resistivity starts to increase slightly. Comparable transition temperatures have been also reported in  other charge-ordered insulators such as $\beta^{\prime\prime}$-(BEDT-TTF)$_{3}$[(H$_{3}$O)Ga(C$_{2}$O$_{4}$)$_{3}$]C$_{6}$H$_{5}$NO$_{2}$ \cite{Yamamoto04},  $\beta^{\prime\prime}$-(BEDT-TTF)$_{4}$(ReO$_{4}$)$_{2}$ \cite{Ihara16} and $\theta$-(BEDT-TTF)$_{2}$\-RbZn(SCN) \cite{HMori98}. As the temperature drops below 72~K, $\rho(T)$ exhibits a steep jump and rises dramatically to a value of $10^7~\Omega$cm at $T=20$~K. The inflection point identified around 50~K maybe related to ordering of the ethylene groups. The strong hysteresis observed for the cooling and warming cycles indicates that a structure distortion is involved in the metal-insulator transition. The transition temperature is best defined by the peak in the logarithmic derivative $-{\rm d}\,\ln\rho/{\rm d}T$ {\rm versus} $T^{-1}$ presented in the inset of Fig.~\ref{fig:dc}(b).
From the Arrhenius plot $\rho(T)\propto \exp\{\Delta_\rho/k_B T\}$ [Fig.~\ref{fig:dc}(b)] the activation energy is estimated to be around $\Delta_\rho = 60$~meV below $T=60$~K and $\Delta_\rho = 170$~meV in the temperature range from 66 to 72~K; for the deuterated and hydrogenated salts an activation energy of 170~meV was previously reported \cite{Lyubovskaya95}.

The lattice constants and volume of the deuterated sibling are expanded compared to \bhgcl; correspondingly the temperature of the sharp metal-insulator transition increases to $T_{CO}=86$~K \cite{Dyachenko95}. Via the interaction between the anion and the terminating ethylene groups of the BEDT-TTF molecules,
deuteration basically works as negative pressure. This implies an enhancement of the effective electron-electron correlation in \dhgcl\
%$\beta^{\prime\prime}$-(D$_8$-BEDT-TTF)$_{2}$\-Hg(SCN)$_{2}$Cl
as the transfer integrals are smaller. In case of one-dimensional organic conductor
it was noted previously that the donor-anion interaction is important for establishing charge order \cite{Pouget15}.

\subsection{Magnetic Properties}
\label{sec:magnetic}
\begin{figure}
	\centering
	\includegraphics[width=0.9\columnwidth]{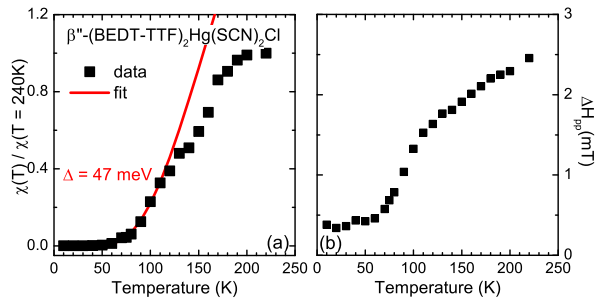}
	\caption{(Color online)(a) Temperature dependence of the relative spin susceptivility of \bhgcl. The solid red line represents a fit of $ \chi(T)$ to Bulaevskii's model of excitations across a singlet-triplet gap $\Delta_\sigma = 47$~meV. (b) Temperature dependence of the linewidth $\Delta H_{pp}$ of the salt. }
	\label{fig:ESR}
\end{figure}

In order to advance our understand of the metal-insulator transition in \bhgcl,
we performed ESR measurements as a function of temperature with the external magnetic field oriented parallel to the conducting ($ab$)-plane.
In Fig.~\ref{fig:ESR} the temperature evolution of the spin susceptibility $\chi(T)$ and linewidth $\Delta H(T)$ are plotted as a function of temperature; the spin susceptibility data are normalized to $\chi(T=240~{\rm K})$. Upon cooling the spin susceptibility drops slightly at elevated temperatures. Around $T=150$~K the decrease becomes more rapid indicating a semi-metallic behavior with a gradual opening of a gap in accord to the minimum in resistance $\rho(T)$, illustrated in the inset of Fig.~\ref{fig:dc}(a).
As the temperature is lowered even further, a non-magnetic insulating state is entered and the spin susceptibility vanishes rapidly as $T\rightarrow 0$. The temperature dependence of linewidth is plotted in Fig.~\ref{fig:ESR}(b); it continuously decreases upon cooling with a pronounced drop around the 120~K and a saturation for $T<T_{\rm CO}$.
It is consistent with other charge ordered  $\beta^{\prime\prime}$-compounds showing non-magnetic ground state \cite{Kawamoto11,Schlueter01,Ward00,CARNEIRO84}.

It is interesting to compare the behavior with the data for \khgcl, where an anti-ferromagnetic insulating ground state was observed \cite{Yasin12}. In the present case a spin-singlet state is realized at low temperatures, and the data $\chi(T)$ can be described with a one-dimensional singlet-triplet picture. From the corresponding Bulaevskii's model \cite{Bulaevskii88}, we expect for the temperature dependence of the spin susceptibility \cite{Dumm00,Dumm11}:
 \begin{equation}
\chi(T)\propto \frac{1}{T} \exp\left\{\frac{-\Delta_\sigma}{k_{B}T}\right\} \quad ;
\label{eq:Bulaevskii}
 \end{equation}
here $\Delta_{\sigma}$ is the static spin gap.
From Fig.~\ref{fig:ESR} we can see that this simple model fits the experimental data of \bhgcl\ quite well
with a spin gap of $\Delta_\sigma = 47$~meV, corresponding to $\Delta_\sigma/k_B = 550$~K.
It is interesting to compare this value with the charge gap of similar size derived from the activated behavior of the electronic transport.

One should note that the spin susceptibility does not show a significant change at $T_{\rm CO}$; the behavior of Eq.~(\ref{eq:Bulaevskii}) extends smoothly well above 100~K. This provides strong evidence for the spin pairing setting in already around 150~K. We will come to similar conclusions from the analysis of our vibrational investigations in Sec.~\ref{sec:vibrational}

\subsection{Optical Properties}
\label{sec:reflectivity}

\begin{figure}
	\centering
	\includegraphics[width=0.8\columnwidth]{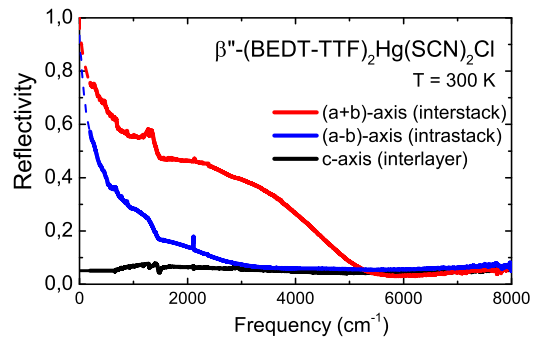}
	\caption{(Color online) Optical reflectivity of \bhgcl\ measured at room temperature with light polarized along all three crystallographic axes as indicated. The extrapolations used below 200~\cm\ are indicated by dashed lines.  }
	\label{fig:ref}
\end{figure}

Fig.~\ref{fig:ref} displays the normal-incidence optical reflectivity of \bhgcl\ as obtained at room-temperature with light polarized along the three principal directions. Perpendicular to the layers ($E \parallel c$) the reflectance is rather low and basically independent on frequency, resembling an insulator; the vibrational features around 1400~\cm\ will be discussed later in detail (Sec.~\ref{sec:vibrational}).

\begin{figure*}
	\centering
	\includegraphics[width=0.8\textwidth]{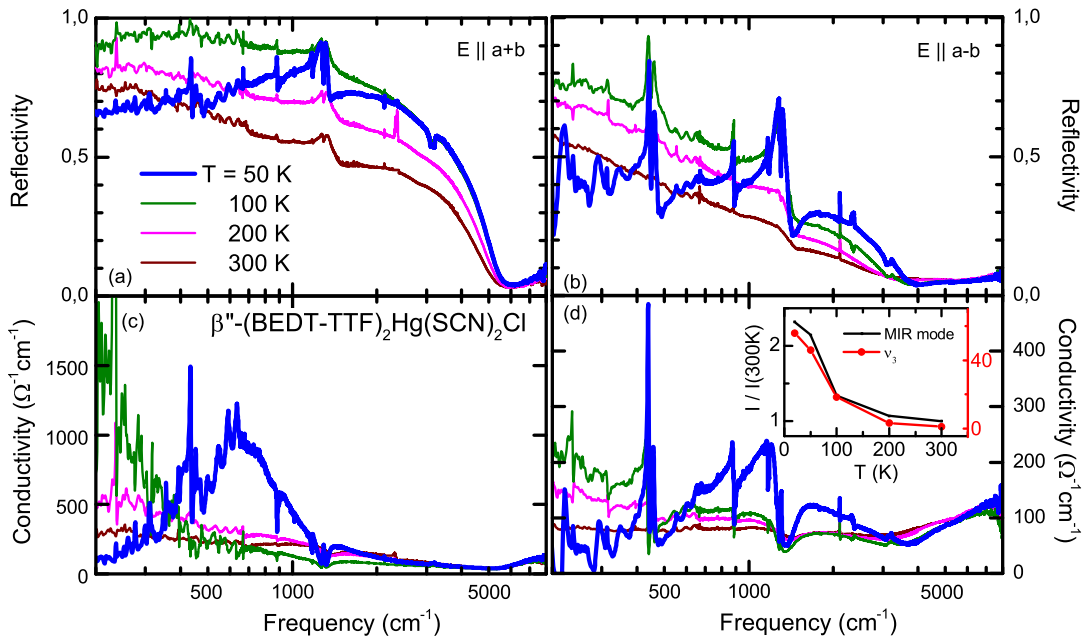}
	\caption{(Color online) (a,b) Optical reflectivity and (c,d) corresponding conductivity spectra of \bhgcl\ measured at different temperatures between $T = 300$ and 50~K for both polarizations perpendicular and parallel to the stacks, i.e.\ $E \parallel (a+b)$ and $E \parallel (a-b)$  \cite{remark2}.
Note the metal-insulator phase transition at $T_{\rm CO}=72$~K leads to a drastic change in the optical properties: the low-frequency reflectivity drops significantly and the Drude component vanishes for $T=50$~K data (thick blue curves).
%The red dashed line in panel (d) illustrates the shift of the mid-infrared mode at around 1200~\cm.
In the inset the intensity evolution of the 1200~\cm\ mode is plotted in comparison to the frequency shift of the vibronic $ \nu_{3}$(A$_g$) mode as temperature is reduced.
}
	\label{fig:cond}
\end{figure*}

Within the highly conducting layer a metallic reflectance  is observed, which extra\-polates to $R=1$ for $\omega \rightarrow 0$ at elevated temperatures.
However, at $T=300$~K only for the polarization $E\perp {\rm stacks}$ a reflection edge can be identified around 5000~\cm.
The overall behavior resembles the
spectra obtained for other BEDT-TTF salts with $\alpha$-, $\beta^{\prime\prime}$- or $\theta$-stacking pattern \cite{Dressel04}. The optical response is rather anisotropic within the quasi-two-dimensional conducting plane over a wide frequency range. As typical for these non-dimerized BEDT-TTF compounds, the reflectivity in the perpendicular direction is significantly larger than $R(\nu)$ measured
along the stacks. This agrees with calculations of the electronic orbitals and bandstructure \cite{TMori98} indicating that the transfer integrals between nearest-neighbor BEDT-TTF molecules are much larger for the interstack direction compared to the direction along the stacks.

Apart from the electronic contributions, several vibrational features can be identified in the mid-infrared spectral region. When probed within the conducting plane, these are attributed to the totally symmetric A$ _{g} $ vibrational modes of the BEDT-TTF molecules coupled with electronic excitations through the electron-molecular vibration (emv) interaction \cite{Girlando86,Yartsev90,Girlando11}.
The sharp peak at approximately 2100~\cm\ is the CN stretching mode in the (SCN)$^{-1}$ entity
of the anion layers \cite{Drichko14}.
It does not change significantly with temperature.

\subsubsection{Electronic contributions}
\label{sec:electronic}

The temperature-dependent reflectivity is plotted in Fig.~\ref{fig:cond}(a,b) for a wide frequency range.
With lowering $T$ a clear plasma edge develops for both polarizations, but at significantly different
frequencies, expressing the in-plane anisotropy: for $E \parallel {rm stacks}$  at 3500~\cm\ and at around 5000~\cm\ perpendicular to it. The position of the edge shifts to higher frequencies upon cooling
because the charge carrier density increase as the lattice contracts.
The overall reflectivity remains much lower than expected for a typical metal;
in particular for the stacking axis, $R(\nu)$ does not exceed 0.5 at $\nu=1000$~\cm\ for $T>100$~K.
Such a behavior is commonly observed for organic conductors \cite{Dressel04}
reflecting the low carrier density and the influence of electronic correlations. Nevertheless, for both orientations the reflectivity increases as the temperature is reduced until the metal-insulator transition takes place around $T_{\rm CO} = 72$~K. At lower temperatures the metallic properties are lost: $R(\nu)$ substantially drops for $\nu < 1000$~\cm\ in both polarizations. Since the crystal becomes transparent at $T=20$~K, multireflection within the crystal leads to interference fringes that hamper a further analysis; thus we constrain ourselves to the $T=50$~K data \cite{remark2}.

A better understanding of the electronic properties can be reached by looking at the temperature-dependent optical conductivity derived via the Kramers-Kronig analysis of the reflectivity spectra and displayed in Fig.~\ref{fig:cond}(c,d). For both polarizations a rather weak Drude-like contribution is observed at room temperature, together with a comparably wide feature in the mid-infrared region centered  between 1000 and 2500~\cm.
When cooling down to $T=100$~K, the spectral weight shifts to low frequencies and a narrow zero-frequency peak appears indicating a weakening of correlations. This tendency is most pronounced for the high-conductive polarization $E \parallel (a+b)$; parallel to the stacks the spectral weight transferred from the higher energy range (up to 1~eV) piles up in the far-infrared as well as in the mid-infrared range around 1200~\cm.

Similar observations have been reported for other organic compounds with quarter-filled conduction bands \cite{Dressel94,Drichko06,Kaiser10,Hashimoto14} and were theoretically described by Merino {\it et al.} \cite{Merino03,Merino05}.
According to their calculations of the extended Hubbard model, the charge carriers become increasingly localized as the effective nearest-neighbor Coulomb repulsion $V/t$ gets important; in addition to the Drude-like term, a finite-energy mode develops in $\sigma(\omega)$ due to charge-order fluctuations and shifts to higher frequencies. Driven by electronic correlations a metal insulator transition eventually takes place: the Drude component vanishes and a mid-infrared excitations  remain at frequencies comparable to the intersite Coloumb repulsion $V$.
%In other words, the mid-infrared term is a measure of correlations.
It is interesting to compare the behavior with the optical spectra of Mott insulators
frequently present in half-filled systems \cite{Faltermeier07,Merino08,Dumm09,Ferber14}
where excitations between the lower and upper Hubbard band allow to determine the Coulomb repulsion $U$. In the latter case, the bands are centered around 2500~\cm\ and possess a typical bandwidth of approximately 0.5~eV \cite{Pustogow17}.

In our spectra taken on \bhgcl\ for $E$ perpendicular to the stacks we observe a very strong band at 500-700~\cm,
resembling the charge fluctuation mode seen in the all-organic superconductor
$\beta^{\prime\prime}$-(BEDT-TTF)$_2$\-SF$_5$\-CH$_2$\-CF$_2$SO$_3$ \cite{Kaiser10}.
Along the stacks the optical conductivity looks rather different and the main feature develops around 1200~\cm, but is present already above the metal-insulator transition.
Although the actual shape of this mode is distorted by the $\nu_3$ antiresonance,
in the inset of Fig.~\ref{fig:cond}(d) we compare the intensity of this 1200~\cm\ band
to the temperature evolution of the $\nu_{3}({\rm A}_g)$ mode.
This similarity in the $T$ dependence implies that the infrared band along the $a$-axis  is related to the emv-coupled $\nu_{3}$ vibration.
It becomes activated because the structure dimerizes along the stacking direction.
The peak lies at much lower position compared to the strongly dimerized $\kappa$-(BEDT-TTF)$_{2}X$ compounds \cite{Dumm09}.
Such a low-lying dimer peak was also reported for  $\beta$-(BEDT-TTF)$_{2}$ICl$_{2}$
where the application of pressure shifts the mode below the Hubbard band  \cite{Hashimoto15}.

From the large anisotropy of the electronic spectra we conclude that stripes of charge-poor and charge-rich molecules likely develop along the ($a+b$) direction where the orbital overlap is largest; they alternate along the stacking direction. Such a large anisotropy has also been observed in other organic conductors such as $\alpha$- \cite{Ivek11}, $\theta$- \cite{Wang01} and $\kappa$-BEDT-TTF salts \cite{Drichko14} with charge-order stripe in that orientation. This is supported by calculations using the extended Hubbard model with an anisotropic intersite Coulomb interaction $V$, which favors stripe-type charge order by a smaller gap, rather than checkboard charge order present in a square lattice \cite{Merino05}. Charge excitations within these one-dimensional channels lead to a band around 600~\cm\ very similar to theoretical predictions \cite{Merino03,Merino05} and previous observations \cite{Kaiser10}.

Below the phase transition at $T_{\rm CO}= 72$~K, the low-energy parts of $R(\nu)$ and $\sigma(\nu)$
alter drastically. A gap opens in the conductivity spectra around 400~\cm\ and 300~\cm\ for the ($a-b$) and ($a+b$) axis, respectively; we estimate $2\Delta_\omega\approx 40-50$~meV, corresponding to approximately 500~K.
The agreement with the activation energy determined from dc resistivity $\rho(T)$ shown in Fig.~\ref{fig:dc} is remarkable.
These values are slightly larger compared to what is expected from mean-field ratio theory:
$2\Delta = 3.53 k_B T_{\rm CO}$. In this regard our findings are in line with other charge-ordered insulators,
such as $\alpha$-(BEDT-TTF)$_{2}$I$_{3}$ ($T_{\rm CO} = 135$~K, $2\Delta_\omega/hc \approx 600$~\cm, \cite{Dressel94,Ivek11}),
but smaller than what is reported for $ \theta$-(BEDT-TTF)$_{2}$RbZn(SCN)$_{4}$: $T_{\rm CO} = 190$~K, $2\Delta_\omega/hc = 300$~\cm\ \cite{Wang01}.

%In many charge-transfer compounds \cite{Nakamura07,Mori98,Kimura05,Yamamoto04,Ihara16} it was shown that charge-order instability takes place hand in hand with a displacement of the molecules or structural changes due to strong coupling with the lattice degree of freedom.
%Theorefore, similar principle can apply to the organic salt studied above. Below the sharp metal-insulator transition, strong modification of lattice occur for both axises resulting in an change of the ration V/t, and push the system to the Charge order state. Combined with the ESR data, we expect that the spin first localized along the stack, and then it very likely pairs together with an opposite spin along the interstack forming singlet dimer state below 75K.

%N_{\text{eff}}(\omega)=\dfrac{2m_{e}V}{\pi e^{2}}\int_{0}^{\omega_{c}}\sigma(\omega)d\omega ,
%\end{equation}

%where m$ _{e} $ is free-electron mass, and V is the volume per formula unit
%Within tight-binding approximation for the one-dimensional organic conductors assuming non-interacting electrons, the carrier density is related to the transfer integral t:
 %\begin{equation}
%N_{\text{eff}}\propto td^{2}\text{sin}\{{\frac{\pi}{2}\rho}\},
%\end{equation}
%Where d is the intermolecular distance, and $ \rho $ is number of electrons per site. For an ideal metal, the

\subsubsection{Vibrational features}
\label{sec:vibrational}

\begin{figure}
	\centering
	\includegraphics[width=0.8\columnwidth]{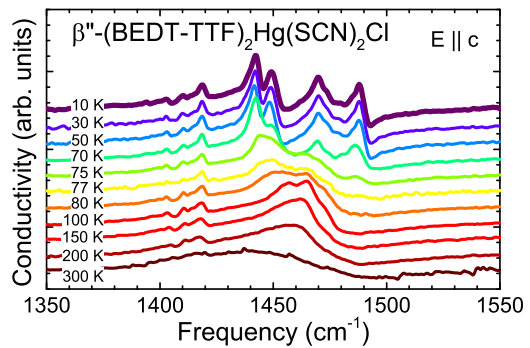}
	\caption{(Color online) Optical conductivity of \bhgcl\ for $E\parallel c$ at different temperatures; the spectra are offset for clarity reasons. Most important is the splitting of the $\nu_{27}({\rm B}_{1u})$ vibrational mode
below $T_{\rm CO}=72$~K, indicating a pronounced charge-ordered state that evolves with temperature.}
	\label{fig:nu27a}
\end{figure}

The infrared-active molecular vibrational mode $\nu_{27}({\rm B}_{1u})$
mode is regarded as the best probe of the local charge on the molecules \cite{Dressel04,Girlando11,Yue10}.
Figure \ref{fig:nu27a} displays the temperature dependence of the optical conductivity spectra measured with the electric field $E||c$, i.e. polarized perpendicular to the conducting layer.
Several {\it ungerade}, infrared-active vibrational B$_{u}$ modes are well resolved in the frequency region between 1300 and 1600~\cm, which is free from disturbing electronic background.
Three weak bands show up at around 1400 to 1425~\cm\ and exhibit almost no temperature dependence;
they can be assigned to the $\nu_{28}({\rm B}_{1u})$ vibration associated with the CH$_{2}$ bending modes \cite{Sedlmeier12}.

The most pronounced feature at room temperature is a very broad band around 1445~\cm\ that corresponds to the
$\nu_{27}({\rm B}_{1u})$ band mainly involving the C=C vibrations.
Such a large linewidth can be explained by dynamical charge-order fluctuations, which also have been discussed and reported in other $\beta^{\prime\prime}$ compounds \cite{Girlando14,Yamamoto04,Yamamoto06,Yamamoto08,Kaiser10}.
Upon cooling the single mode becomes narrower, shifts up in frequency and develops a double peak structure around $T=100$~K. This is not a big surprise considering the fact that two distinct types of BEDT-TTF molecules (A, B) reside in the unit cell, as illustrated in Fig.~\ref{fig:structure}(b).
As the temperature is reduced below the phase transition at 72~K,  the feature splits into four well distinct peaks
at 1442.2, 1449.0, 1469.8 and 1487.6~\cm, giving evidence that the insulating state exhibits charge disproportionation, i.e.\ there are four distinct types of BEDT-TTF molecules.

\begin{figure}
	\centering
	\includegraphics[width=0.9\columnwidth]{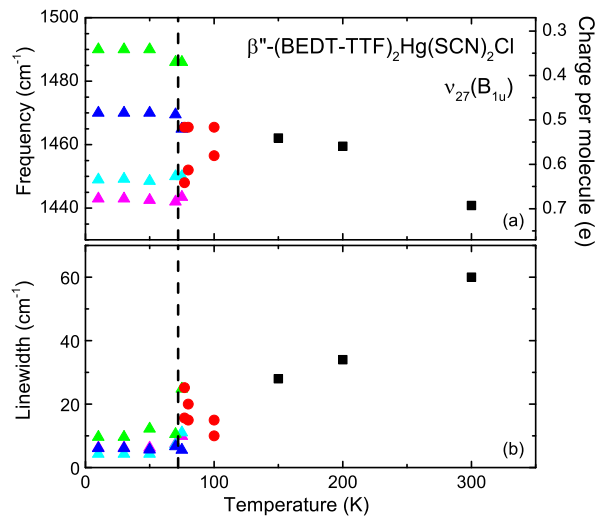}
	\caption{(Color online) (a)~Frequencies of the $\nu_{27}({\rm B}_{1u})$  modes plotted against temperature. The scale on the right axis corresponds to the charge per site according to Eq.~(\ref{eq:splitting}). (b)~Temperature dependence of the linewidth of the $\nu_{27}$. Above $T=100$~K a single mode is observed (black squares);
in the fluctuation regime ($72~{\rm K}<T<100$~K) two modes are fitted and indicated by red dots. Triangle with different colors are used to identify the four modes in the charge-ordered state. The dashed line indicates the metal insulator transition $T_{\rm CO}=72$~K also observed in dc resistivity $\rho(T)$.}
	\label{fig:splitting}
\end{figure}
To quantitatively characterize the observed charge imbalance, we fit the spectra of Fig.~\ref{fig:nu27a} by simple Lorentz oscillators. The temperature profile of the center frequency and linewidth at half maximum is displayed
in Fig.~\ref{fig:splitting}. The charge per molecule can be evaluated from the vibrational frequency by using the relationship \cite{Yamamoto05,Girlando11}:
 \begin{equation}
\nu_{27}(\rho)=1398\,\text{cm}^{-1}+140(1-\rho)\,\text{cm}^{-1}/e \quad ,
\label{eq:splitting}
\end{equation}
where $\rho $ is the site charge in units of the elementary charge $e$; the corresponding scale is indicated on the right axis of Fig.~\ref{fig:splitting}(a).
The $\nu_{27}({\rm B}_{1u})$  mode exhibits a slight blue shift from 1462 to 1470~\cm\ as the lattice hardens when cooling down in the metallic state; at $T=100$~K the peak approaches the $\rho = 0.5e$ value.
The vibrational feature concomitantly narrows down from 60 to 15~\cm\ because thermal effects diminish. As the phase transition is approached, in the temperature range between 100 and 75~K, the molecular vibrations split in two peaks with the separation growing from approximately 9~\cm\  continuously up to 18~\cm. As the phase transition is approached, the lines become rather broad pointing towards fluctuation effects.
A similar broadening in the linewidth was observed in other charge-ordered compounds and related to charge fluctuations already present in the metallic state \cite{Yamamoto04,Yamamoto06}.
\begin{figure}
	\centering
	\includegraphics[width=0.8\columnwidth]{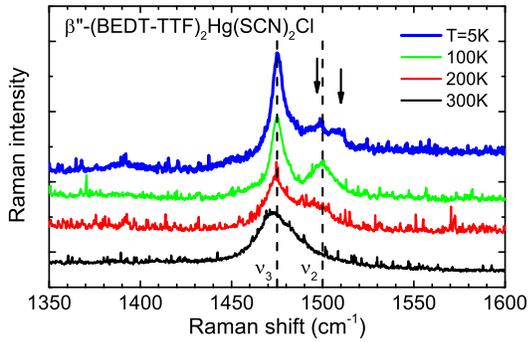}
	\caption{(Color online) Raman shift for the $\nu_{2}$ and $\nu_{3}$ modes of \bhgcl\ at several selected temperatures as indicated. Black arrows indicate the splitting of the $\nu_{2}$ vibration in the charge-ordered state. }
	\label{fig:Raman}
\end{figure}

\begin{figure*}
\centering
\includegraphics[width=0.8\textwidth]{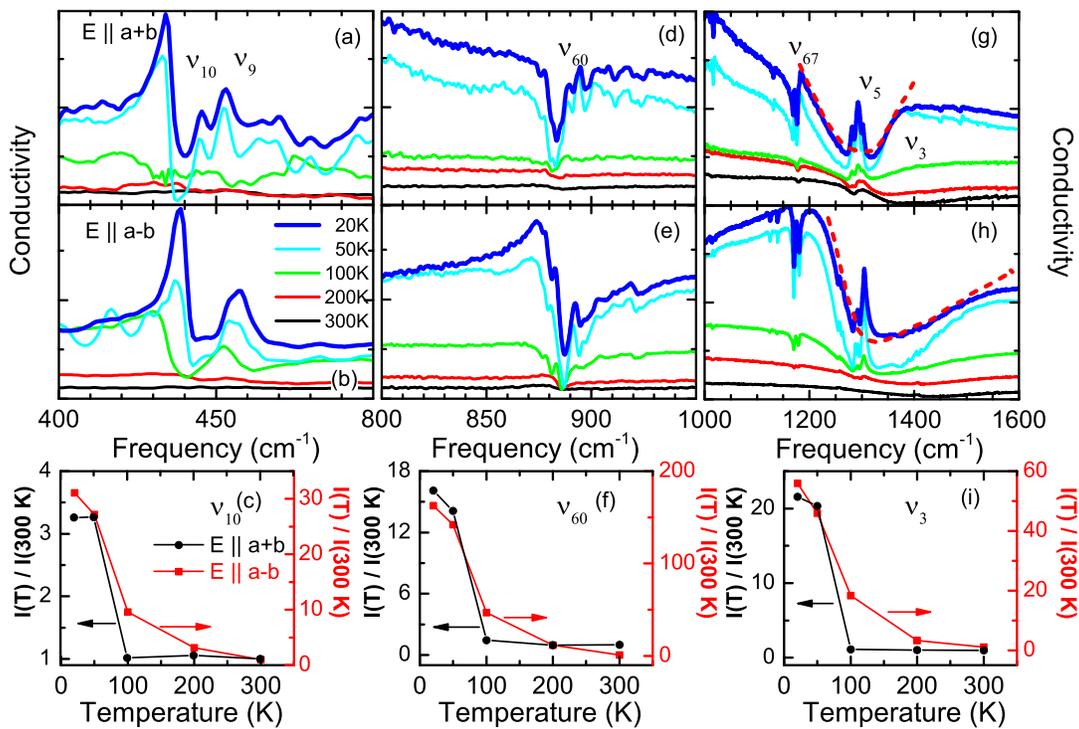}
\caption{(Color online) Temperature dependence of the emv-activated vibrational modes in \bhgcl\ probed within the conducting $(ab)$-plane. (a,b)~The $\nu_{9}({\rm A}_{g})$ and $\nu_{10}({\rm A}_{g})$ modes show up only in the charge-ordered state and appear more diverse for the  polarization (a) perpendicular to the stacks (\textcolor{red} [$E\parallel (a+b)$] compared to (b) the stacking direction [$E\parallel (a-b)$.]
(d,e)~The charge sensitive $\nu_{60}({\rm B}_{3u})$ vibration comes as a strong dip for perpendicular polarization (panel d) but exhibits a Fano shape along the stacks (panel e). (g,h)~In the range between 1000 and 1400~\cm\ the emv-coupled fully-symmetric $\nu_3({\rm A}_g)$ vibration shows up as a broad dip indicated by red dashed line. The narrow peak around 1300~\cm\ is assigned to the $\nu_{5}({\rm A}_g)$ CH$_2$ wagging mode. Also seen is the $\nu_{67}({\rm B}_{3u})$ as a strong dip at 1175~\cm\ in both polarizations. The curves are shifted for clarity.
In panels (c,f,i) exhibit the temperature evolution of the $\nu_{10}$, $\nu_{60}$ and  $\nu_{3}$ modes, respectively, where the left axes (black) correspond to the polarization perpendicular to the stacks and the right axes (red) to the parallel polarization.
\label{fig:vibronic}
}
\end{figure*}

When the resistivity data give evidence of a metal-insulator phase transition at $T_{\rm CO}=72$~K also the molecular vibrational modes exhibit a clear splitting in four distinct lines, suggesting the existence of four non-equivalent BEDT-TTF molecules in the unit cell and the breaking of inversion symmetry. The charge difference of the four modes remains temperature independent. This behavior follows the general trend observed in two-dimensional charge-ordered systems, such as $\alpha$-(BEDT-TTF)$_{2}$I$_{3}$ \cite{Yue10,Ivek11,Beyer16} and \khgcl\ \cite{Drichko14,Ivek17b}.
Based on Eq.~(\ref{eq:splitting}) the lower-frequency modes correspond to the $+0.68e$ and $+0.64e$ charges on the BEDT-TTF molecule,  and the upper frequency ones to $+0.49e$ and $+0.34e$  for $T=10$~K.
The maximum charge disproportion $2\delta_\rho =0.34e$ is about half the value found in $\alpha$-(BEDT-TTF)$_{2}$I$_{3}$, similar to the imbalance observed for $\beta^{\prime\prime}$-(BEDT-TTF)$_{4}$$M$(CN)$_{4}\cdot$H$_{2}$O ($M$ = Ni, Pd),  but lager than those reported for the polymorph \khgcl\ \cite{Yamamoto08,Drichko14} %,Ivek17b}
and the charge-fluctuating superconductor $\beta^{\prime\prime}$-(BEDT-TTF)$_{2}$SF$_{5}$CH$_{2}$CF$_{2}$SO$_{3}$ \cite{Kaiser10}.

The charge order pattern forms in such a way that the total energy of the system is
minimized; in a first approximation we can neglect the transfer integral and discuss the intersite Coulomb repulsion only. For the $\beta^{\prime\prime}$-BEDT-TTF salts, similar to $\theta$- and $\alpha$-phase crystals,
the intermolecular distances are significantly shorter along the stacking axis.
Consequently the intermolecular Coulomb repulsion is more pronounced in this direction compared to the perpendicular orientation \cite{MoriBook,Mori00}.
This results in a charge alternation along the ($a-b$)-axis: A\,a\,B\,b\,A\,a\,B\,b, denoting the A-type and B-type molecules shown in Fig.~\ref{fig:structure}(b) with capital and small characters referring to the charge rich and charge poor sites;
the neighboring stacks will be arranged like B\,b\,A\,a\,B\,b\,A\,a, leading to horizontal stripes.
T.Yamamoto {\it et al.} classified some of the charge-ordered BEDT-TTF  salts
and suggested a phase diagram for the $\beta^{\prime\prime}$-type compounds \cite{Yamamoto08}.
The amount of charge disproportionation and the alternation of short and long intermolecular bonds
along the stacks puts the \bhgcl\ into group I.
Going beyond purely electronic models Mazumdar {\it et al.} suggested alternative patterns for the $\alpha$-phase materials \cite{Mazumdar99,Mazumdar00} that might also be applicable to  the present compound; however all experimental results on $\alpha$-(BEDT-TTF)$_{2}$I$_{3}$, such as infrared spectroscopy \cite{Ivek11}, NMR \cite{Ishikawa16} and x-ray studies \cite{Kakiuchi07}, evidence a horizontal stripe structure. Thus we concluded that the charge-ordered state in \bhgcl\ develops horizontal stripes.

Apart from the infrared-active $\nu_{27}({\rm B}_{1u})$ mode, there are two fully symmetric C=C stretching vibrations that allow to determine the charge on the BEDT-TTF molecules with the help of Raman spectroscopy.
In Fig.~\ref{fig:Raman} the shift of Raman-active $\nu_{2}({\rm A}_g)$ and $\nu_{3}({\rm A}_g)$ modes
is displayed for several selected temperatures.
At room temperature only one asymmetric band is detected at 1472~\cm\ with a hump at round 1500~\cm.
As the temperature is reduced, the 1472~\cm\ peak gets narrow without any shift or splitting, while the 1500~\cm\
feature increases in intensity and splits into two modes as $T=10$~K is approached.
We assign the 1472~\cm\ feature to the $\nu_{3}$ molecular vibration, which is insensitive to the charge
and strongly couples to the electronic background as shown by the strong infrared intensity \cite{Yamamoto02,Yamamoto04,Yamamoto06,Yamamoto08} and discussed in more detail below.
The right band at 1500~\cm\ corresponds to the $\nu_{2}$ mode, which is weakly coupled to electronic background and suitable for estimating the ratio of charge disproportionation. From the splitting of $\nu_{2}({\rm A}_g)$ feature we calculate $2\delta_\rho = 0.2e$ in the charge-ordered state according to \cite{Yamamoto05,Girlando11}:
 \begin{equation}
\nu_{2}(\rho)=1447\,\text{cm}^{-1}+120(1-\rho)\,\text{cm}^{-1}/e \quad .
\end{equation}
This value is smaller than the charge imbalance estimated from the $\nu_{27}({\rm B}_{1u})$ splitting observed by infrared spectroscopy (Fig.~\ref{fig:splitting}). We cannot rule out that some minor peaks could not be resolved due to the poor signal to noise level in our Raman signal.

Several molecular vibrational modes show up rather prominently in the infrared spectra of \bhgcl\ measured within the ($ab$)-plane due to strong coupling to the charge transfer band. These modes are charge sensitive but are also susceptible to structural changes.
The  $\nu_{10}({\rm A}_{g})$ and  $\nu_{9}({\rm A}_{g})$ modes observed in the far-infrared range involve C-S streching vibrations. As shown in Fig.~\ref{fig:vibronic}(a) for $E$ perpendicular to the stacks, the two modes can hardly be detected in the metallic state due to  screening effects;
but when entering the charge-ordered insulating state several strong peaks can be identified. Within the plane, for $E\parallel {\rm stacks}$ [Fig.~\ref{fig:vibronic}(b)], we always detect two modes around 440 and 460~\cm, assigned to $\nu_{10}$ and $\nu_{9}$, respectively. Upon cooling the two features shift to higher frequencies, become sharper and more intense, as plotted in Fig.~\ref{fig:vibronic}(c).
A similar multiple splitting of the $\nu_{9}$ and $\nu_{10}$ modes in the charge-ordered state was reported for other organic BEDT-TTF compounds \cite{Drichko06,Ivek11}.

Around $\nu=880$~\cm\ for both polarization directions we identify the $\nu_{60}({\rm B}_{3u})$ mode that involves ring-breathing vibration [Fig.~\ref{fig:vibronic}(d,e)];
this molecular vibration is known to be very sensitive to charge disproportion and dimerization \cite{Musfeldt05}.
The mode splits into several dips for both polarizations at low temperatures;
for $E\parallel {\rm stacks}$ the intensity strongly increases with lowering the temperature, as
displayed in Fig.~\ref{fig:vibronic}(f). Interestingly in the case of dimerized Mott insulators, basically
no significant temperature dependence was observed \cite{Sedlmeier12}.

The broad dip at around 1300~\cm\ shown in Fig.~\ref{fig:vibronic}(g,h) is assigned as the fully symmetric
$\nu_{3}$ vibration that has the strongest emv-coupling constant among all the A$_{g}$ modes and exhibits a down-sift in frequency over 100~\cm\ compared to 1450~\cm\ line measured by Raman spectroscopy (Fig.~\ref{fig:Raman}).
There is a rather narrow three-peak structure at 1282, 1293, and 1302~\cm\
related to the $\nu_5({\rm A}_g)$ mode
present in both orientations within the ($ab$)-plane, which indicates that the symmetry of the unit cell is broken at low temperatures.
For the polarization parallel to the stacks the intensity of the $\nu_3$ mode increases progressively as the temperature decrease, while in the perpendicular directions a strong growth is observed only below the transition temperature $T_{\rm CO}$.
By comparing the conductivity spectra of \bhgcl\ in this range with those reported for strongly dimerized
$\kappa$-BEDT-TTF salts \cite{Faltermeier07,Dumm09,Sedlmeier12}, we conclude that the intra-dimer charge-transfer band is located much closer to the vibrational modes. In the case of \etcl\ and \etcn\ the bands appear around 3000~\cm\ and consequently the emv coupled vibrations show up as Fano-like features.
In line with calculations based on a one-dimensional dimerized tight-binding model \cite{Bozio87},
we suggest that the $\beta^{\prime\prime}$-compound studied here is slightly dimerized.
From the gradual enhancement of the vibrational intensity with decreasing temperature for $E\parallel {\rm stacks}$,
plotted in Fig.~\ref{fig:vibronic}(c,f,i), we conclude that the dimerization is more pronounced along the stacks than in the direction perpendicular  to it.
From the analysis of the temperature behavior evolution of the pure vibrational modes and the emv-coupled vibronic features, we can identify the low-temperature ground state as a charge-ordered insulating state, where the gradual structural distortion  and charge disproportionation are closely related.
Already below 150~K a continuous phase transition sets in gradually,
the electrical resistivity turns from a metallic to an insulating behavior, until the insulating charge-ordered state is finalized at $T_{\rm CO}=72$~K and $\rho(T)$ shoots up.
In Sec.~\ref{sec:magnetic} we have drawn the same conclusion from the temperature dependence of the ESR spectra.
\\

\section{Conclusions}

From our comprehensive characterization and intense optical study we conclude that \bhgcl\
enters a charge-ordered insulating state at $T_{\rm CO}=72$~K  where pronounced charge disproportionation occurs with $2\delta_\rho=0.34$ obtained from infrared vibrational spectroscopy. The charge-rich molecules arrange in stripes perpendicular to the stacks, i.e. along ($a+b$)-direction.
The dc resistivity yields an activated behavior at low temperatures with an energy gap of approximately $\Delta_{\rho}=60$~meV, that agrees well with the gap observed in the optical spectra.
The charge imbalance starts to develop in the temperature range $72~{\rm K}<T<100$~K; at elevated temperatures ($T>100$~K) all the way up to room temperature, charge fluctuations can be identified.
Around 150~K the metallic behavior turn into a semiconducting temperature dependence of the resistivity. In this temperature range  lattice deformations along the BEDT-TTF stack are detected by the simultaneous enhancement of the emv-coupled vibronic mode and dimerization excitation for $E \parallel {\rm stacks}$.
The dimerization also leads to a pairing of the electron spins and a spin-gapped magnetic ground state with $\Delta_{\sigma}=47$~meV obtained from our ESR experiments.
Temperature dependent x-ray studies could further confirm our conclusions.

\begin{acknowledgements}
We thank Gabriele Untereiner for continuous experimental support and Andrej Pustogow and
Mamoun Hemmida for valuable discussions.
We thank Wolfgang Frey for collection of the X-ray data.
The project was supported by the Deutsche Forschungsgemeinschaft (DFG) via DR228/39-1 and DR228/52-1 and by the Deutscher Akademische Austauschdienst (DAAD). The work in Chernogolovka was supported by FASO Russia, \#\ 0089-2014-0036.
\end{acknowledgements}

%\bibliography{reference}

\begin{thebibliography}{63}
\bibitem[{rem({\natexlab{b}})}]{remark2}
\bibinfo{note}{The small interferences observed below 700~\cm are due to the
  multirefelction of the infrared window, which are temperature independent.
  Below $T=20$~K, additional strong interferences appear in the very
  far-infrared region due to the transparence of the thin crystal.}
\end{thebibliography}

\end{document}